\documentclass[%
 reprint,
 amsmath,amssymb,
 aps,
 prl
]{revtex4-2}

\usepackage{graphicx}
\usepackage{dcolumn}
\usepackage{bm}
\usepackage{braket}
\usepackage{amsfonts}
\usepackage{soul}
\usepackage[svgnames,table]{xcolor}

\begin{document}

\preprint{APS/123-QED}

\title{Deep learning for disordered topological insulators through entanglement spectrum}

\author{Alejandro Jos\'e Ur\'ia-\'Alvarez}
 \email{alejandro.uria@uam.es}
 \affiliation{Departamento de F\'isica de la Materia Condensada, Condensed Matter Physics Center (IFIMAC), and Instituto Nicol\'as Cabrera (INC),
Universidad Aut\'onoma de Madrid, Cantoblanco 28049, Spain}
 \author{Daniel Molpeceres-Mingo}
\affiliation{Departamento de F\'isica de la Materia Condensada, Condensed Matter Physics Center (IFIMAC), and Instituto Nicol\'as Cabrera (INC),
Universidad Aut\'onoma de Madrid, Cantoblanco 28049, Spain}
 \author{Juan José Palacios}
 \email{juanjose.palacios@uam.es}
\affiliation{Departamento de F\'isica de la Materia Condensada, Condensed Matter Physics Center (IFIMAC), and Instituto Nicol\'as Cabrera (INC),
Universidad Aut\'onoma de Madrid, Cantoblanco 28049, Spain}

\date{\today}


\begin{abstract}

Calculation of topological invariants for crystalline systems is well understood in reciprocal space,  allowing for the topological classification of a wide spectrum of materials. 
In this work, we present a technique based on the entanglement spectrum, which can be used to identify the hidden topology of systems without translational invariance. By training a neural network to distinguish between trivial and topological phases using the entanglement spectrum obtained from crystalline or weakly disordered phases, we can predict the topological phase diagram for generic disordered systems. This approach becomes particularly useful for gapless systems, while providing a computational speed-up compared to the commonly used Wilson loop technique for gapful situations. Our methodology is illustrated in two-dimensional models based on the Wilson-Dirac lattice Hamiltonian.

\end{abstract}                              

\maketitle

\textit{Introduction}---
Identification of topological materials requires the computation of quantities such as the Berry phase, the Chern number \cite{tknn} or the $\mathbb{Z}_2$ index \cite{z2kanemele}, the so-called topological invariants. The evaluation of Wilson loops  \cite{z2pack}, the most common methodology used to unveil the presence of invariants, works well for crystalline systems since they are performed in reciprocal space, to the point of allowing for high-throughput screening of materials \cite{highthroughputdiscovery, marzari}. Other approaches based on reciprocal space to condensed matter topology exist such as topological quantum chemistry \cite{topquantumchemistry, Vergniory2019}, symmetry indicators \cite{PhysRevB.76.045302, Po2017, Song2018}, or the scattering invariant approach \cite{scatteringinvariant}. Alternatively,
one can resort to the bulk-boundary correspondence: if the system is topological, we expect the presence of conducting surface states. By calculating different observables, such as the conductance or the density of states, one strives to find evidence of the topology, without directly computing any invariant \cite{topologicalmetals}.

The use of the Wilson loop technique requires the existence of a direct band gap everywhere in the Brillouin zone (e.g., no band crossings) near the Fermi level. An overall gap does not need to exist and the system may still be topologically non-trivial. For disordered or non-translationally invariant systems where the bands are not well-defined (e.g. in an open system), or are defined in a very small Brillouin zone (a large supercell  where the spectrum becomes essentially discrete) the direct gap concept is lost. In the latter case, if an overall gap is clearly visible, the Wilson loop can still be calculated \cite{bismutheneamorphous}. However, the absence of an spectral gap does not preclude a non-trivial topology \cite{PhysRevB.85.205136}.


The recent introduction of topological markers such as the Chern marker \cite{localchern} and the Bott index \cite{bottindex, bott_wannier}, has changed our view that working in the reciprocal space is essential since they manage to provide information on the topology of the system based exclusively on real-space computations. 
For time-reversal topological insulators, on the other hand, much less has been reported in this regard, with the exception of a variant of the Bott index which enables a real-space study of the $\mathbb{Z}_2$ number \cite{bottindex} or a spin Bott index \cite{spinbott} in analogy with the spin Chern number \cite{spinchern}.

Some quantities that have been shown to be related to the topology of the system are the entanglement spectrum and entropy \cite{entanglementhaldane, entanglementbernevig, traceindex, entanglementbi, entanglementashvin, entanglementfidkowski, inversionentanglementbernevig, turner2010band, PhysRevLett.110.046806, PhysRevB.88.115114, Wang_2014, PhysRevB.89.195120}. These magnitudes measure the degree of entanglement of our ground state between two halves of the system.  In the presence of translational symmetry, the $\mathbb{Z}_2$ index can be defined from the entanglement spectrum \cite{traceindex} through its flow, much in analogy with the hybrid Wannier charge center (HWCC) flow \cite{wannierrepresentationvanderbilt}. Resorting to the insulating picture, it is known that a topologically non-trivial ground state presents in-gap single-particle states that circulate around the material, wrapping it. The flow of the entanglement spectrum thus reflects the appearance of surface states upon the separation of the two halves.
Again, however, for disordered systems where translational symmetry is lost, we cannot resolve the spectral flow, unless one resorts to inconveniently large supercells. 

Here we will show that one can still use the entanglement spectrum for disordered systems where no momentum component is conserved  with the aid of artificial neural networks (ANNs). Machine learning (ML) algorithms, and ANNs in particular, have been shown to accurately predict topological phases on a wide range on inputs, such as wavefunctions \cite{percolation, PhysRevB.102.054107, PhysRevLett.124.226401}, density matrices \cite{nndensitymatrix, PhysRevB.102.134213}, Berry curvature \cite{ann_curvature}, Hamiltonians \cite{ann_h, PhysRevB.98.085402}. It was also demonstrated that the entanglement spectrum can be used to train ML algorithms to identify topology in translationally invariant systems \cite{PhysRevB.102.054512, PhysRevB.104.165108}, as a function of disorder in one-dimensional AIII models \cite{zhuang2020classification}, or localization phases in interacting systems \cite{PhysRevLett.121.245701}.
In our case, we consider disordered two-dimensional (2D) time-reversal topological insulators. By training an ANN to differentiate between topological and trivial entanglement spectra in models whose invariant is known or can be computed through the Wilson loop technique, we show that we can predict the topology of the system without resorting to the calculation of momentum-space flows. More importantly, our ANN is blind to the absence or existence of a gap in the system.

\textit{Model}---
Our choice is the Wilson-Dirac lattice fermion model, which is a discretized version of the Dirac Hamiltonian on a cubic lattice, with an additional mass term to remove unphysical states when comparing with the continuum Hamiltonian \cite{wilsonchandra}. In real-space, this tight-binding model reads in the following way \cite{montvay}:
\begin{align}
    H = \sum_{i,\mu}&\left[ i\frac{t}{2}c^{\dagger}_{i+\mu}\alpha_{\mu}c_{i} + \frac{r}{2}c^{\dagger}_{i+\mu}\beta c_{i} + \text{h.c.}\right]  \nonumber \\
    &+ (M - 3r)\sum_i c^{\dagger}_i\beta c_i
\label{wilsondiracrealspace}
\end{align}
where the index $i$ sums over lattice positions, and $\mu$ sums over spatial coordinates ($\mu=x,y,z$). In other words, hopping is only considered between first neighbours along the cartesian axis (a cubic lattice is assumed). $\{\alpha_{\mu}\}_{\mu}, \beta$ denote gamma matrices, given by $\alpha_{\mu} = \sigma_{x}\otimes\sigma_{\mu}$, $\beta=\sigma_z\otimes I$.



In the following, we will fix $r=1$ and $t=1$ and we will restrict ourselves to a 2D square lattice.
This model is known to be able to describe topological insulators \cite{fradkin}, with the parameter $M$ tuning the different topological phases. For $1>M>3$ and $3>M>5$, it describes a topological insulator, while for any other value it corresponds to a trivial insulator. This can be readily checked by computing the $\mathbb{Z}_2$ invariant with the help of the HWCC or eigenvalues of the Wilson loop $W(k_y)=\prod_{k_i\in\text{path}}M_{k_i,k_{i+1}}$, where $M_{k_i, k_{i+1}}=U^{\dagger}(k_i, k_y)U(k_{i+1}, k_y)$ and $U(k_i, k_y)$ is the unitary matrix that diagonalizes the Bloch Hamiltonian in the atomic gauge \cite{vanderbilt_book}. 

It has also been shown that this model can also realize non-trivial topology in disordered systems \cite{percolation, amorphous, disorderpotential_wilson, grushin2021topological}. Following \cite{amorphous}, we introduce a generalized version of the Wilson-Dirac model (\ref{wilsondiracrealspace}) to describe crystalline disorder or amorphous solids:
\begin{align}
    H = \sum_{i,j}& \frac{i}{2}t(R)c^{\dagger}_i(\sin\phi\sin\theta\alpha_x + \sin\phi\cos\theta\alpha_y +  \nonumber\\
    &\cos\theta\alpha_z - i\beta)c_j+ \sum_i\beta (M - 3)c^{\dagger}_ic_i.
\end{align}
Here, the variables $(R, \phi, \theta)$ denote the spherical coordinates of the vector determined by the relative position between  lattice sites $i, j$, randomly placed near their original crystal positions. The degree of disorder is characterized by the parameter $\Delta r$, which measures the characteristic distance of the site from its crystal position. As for the hopping amplitude $t\equiv t(R)$, we introduce a dependence with the bond length $R$ through an exponential law:
$$t(R) = \exp\left({\frac{a-R}{a}}\right)\theta_H(R - R_{c}).$$
where $a$ is some reference lattice spacing (which here will be set $a=1$).
Thus, when $R=a$, we recover $t=1$ as in the original model. Note that there is also a Heaviside step function $\theta_H$, which serves as a cut-off for bonds between atoms that are too far apart, $R_c$ being the cut-off distance.
It can be seen that when the lattice sites are restricted to the cubic lattice, we recover the same Hamiltonian in  (\ref{wilsondiracrealspace}). 
\\

\textit{Entanglement}---
Entanglement can provide information about correlations in complex systems, depending on the chosen vector space, such as orbital, spin or spatial \cite{PhysRevLett.110.046806, PhysRevB.88.115114}. In the context of topological materials, it was shown that an entanglement cut of the system into two halves reveals a non-trivial flow of the entanglement spectrum, which can be understood as edge states appearing due to the virtual edge formed by the entanglement cut \cite{entanglementashvin, entanglementfidkowski, inversionentanglementbernevig, traceindex, turner2010band}. Alternatively, it was shown that the entanglement spectrum can be interpreted as coarse-grained hybrid Wannier centers \cite{entanglementwannierarovas, entanglementwannierinterpolation, Lee_2014}, therefore measuring charge pumping across the cut \cite{wannierrepresentationvanderbilt, wilsonloopz2dai}.

The entanglement cut means partitioning our $N$-particle Hilbert space $\mathcal{H}$ into two subspaces $\mathcal{H_{A,B}}$ such that $\mathcal{H}=\mathcal{H_A}\otimes\mathcal{H_B}$. According to the Schmidt decomposition theorem, any state $\ket{\psi}\in\mathcal{H}$ can be written uniquely as $\ket{\psi}=\sum_{i}e^{-E_i}\ket{\alpha_i}\otimes\ket{\beta_i}$, where $\{\ket{\alpha_i}\}_i$, $\{\ket{\beta_i}\}_i$ denote some basis states from the corresponding Hilbert spaces $\mathcal{H}_{A/B}$, and $\{E_i\}_i$ form the many-body entanglement spectrum. From the reduced density matrix, defined as $\rho_A=\mathrm{Tr}_B\rho=e^{-H_e}$, $H_e$ being the entanglement Hamiltonian, we extract the entanglement spectrum as its many-body eigenvalues. 

While in principle we could use this theorem to decompose the ground state, in practice it is better to use an alternative approach. It was proven that for non-interacting systems, the entanglement spectrum can be obtained from the one-particle correlation function \cite{peschel}:
\begin{equation}
    C_{ij} = \braket{\Psi|c^{\dagger}_ic_j|\Psi},\ i,j\in A
\end{equation}
where $\ket{\Psi}=\prod_{n,k}c^{\dagger}_{nk}\ket{0}$ is the Fermi sea.
The eigenvalues of $C$ are called single-particle entanglement spectrum, and they are related to the spectrum of $\rho_A$ via a monotonic function \cite{peschel}, meaning that we can use either indistinctly.
For our purposes, it suffices to use the single-particle entanglement spectrum, although the many-body one can be obtained by filling states in the partitions in all possible combinations \cite{traceindex}.

If we define $H'=\frac{1}{2}-C$, $C$ evaluated over the whole system can be regarded as a flattened Hamiltonian \cite{turner2010band}, with an energy band $\varepsilon=-1/2$ for occupied states, and $\varepsilon=1/2$ for unoccupied states. Then, by virtue of the adiabatic theorem \cite{griffiths}, since both $H$ and $H'$ have a gap, they share the same topology as there is a family of Hamiltonians that interpolate between them \cite{entanglementfidkowski}. Thus, the restriction of $C$ to a half-space results in edge states appearing in its spectrum, connecting the valence and conduction bands.

Following the model we have introduced, the rest of the discussion will be in 2D. In case that our system has translational symmetry, the entanglement cut will preserve it in one direction, meaning that, e.g., $k_y$ is a good quantum number. The one-particle correlation matrix can be calculated as a function of $k_y$:
\begin{equation}
    C_{ij}(k_y)=\braket{\Psi|c^{\dagger}_{ik_y}c_{jk_y}|\Psi}=\sum_{k_x}e^{k_x(i-j)}P_{ij}(\textbf{k})
\end{equation}
allowing to obtain the entanglement spectrum as a function of $k_y$, thus resolving the spectral flow characteristic of the edge states \cite{traceindex}. Here, $P_{ij}(\textbf{k})$ denotes the matrix elements of the projector onto the ground state.

{
From a computational standpoint, the  ${k}$-resolved entanglement spectrum approach does not provide an advantage relative to the computation of HWCC, since both scale as $\mathcal{O}((N\times N)^3\times nk\times npaths)$, where $N$ defines the bidimensional system size.  However, for amorphous or disordered systems, where a supercell is needed, the use of the real-space entanglement spectrum proves to be advantageous as it drastically reduces the number of diagonalizations needed, while still allowing us to determine the $\mathbb{Z}_2$ invariant. Furthermore, its use is mandatory when the systems become gapless, as it usually happens for strong disorder.}
\\

\textit{Neural Network}---
Our aim now is to extract qualitative features from the entanglement spectrum where the topology is expected to be encoded. To be able to tell apart the spectrum of a trivial case from that of a topological one, on top of the automated invariant computation, we introduce an ANN. For topological materials one would naively expect states within the gap of the entanglement spectrum, although flattened Hamiltonians might present midgap energies as well without being necessarily topological \cite{inversionentanglementbernevig}. Otherwise, the spectrum does not present intricate features, meaning that a simple network suffices to classify correctly the phases.
As for the network architecture, our election was a one-dimensional convolutional neural network, with a sigmoid output for binary classification. To tackle the three-dimensional case, the output would have to be a softmax with three neurons, to distinguish between weak and strong topological phases, and trivial.  Training with more complex networks (e.g. more convolutional layers and/or more dense layers) does not seem to improve the performance. One should be aware that inversion-symmetric materials can present non-trivial spectral flows \cite{entanglementashvin}, which could result in false positives. Since we are dealing with disordered topological insulators, this will not be problematic.  

The training of the ANN is done by explicitly computing the $\mathbb{Z}_2$ invariant (when possible) and associating it with its corresponding entanglement spectrum.
We first use data from the crystalline regime 
(zero disorder), whose invariant is easy to compute, and then associate it with the entanglement spectrum corresponding to a supercell big enough so that it is representative of the actual samples for which we want to determine their topology. For completeness, we also use spectra corresponding to the low disorder region ($\Delta r < 0.05$) to train the ANN. The only difficulty here is that the computation of the topological invariant is expensive since a supercell is needed and cannot be done for strong disorder where the gap is lost. Nevertheless, as we show below, non-trivial topology can be predicted beyond the training set.

Two different systems will be studied with this methodology: first, a square lattice which is increasingly deformed by random displacements of the atoms, to the point of the crystal becoming amorphous. Second, a fractal lattice, specifically the Bethe lattice. We will show that in both cases we are able to predict the presence of topological phases by means of the entanglement spectrum, as confirmed by the direct observation of edge states.
\\

\begin{figure}[t]
    \centering
    \includegraphics[width=1\columnwidth]{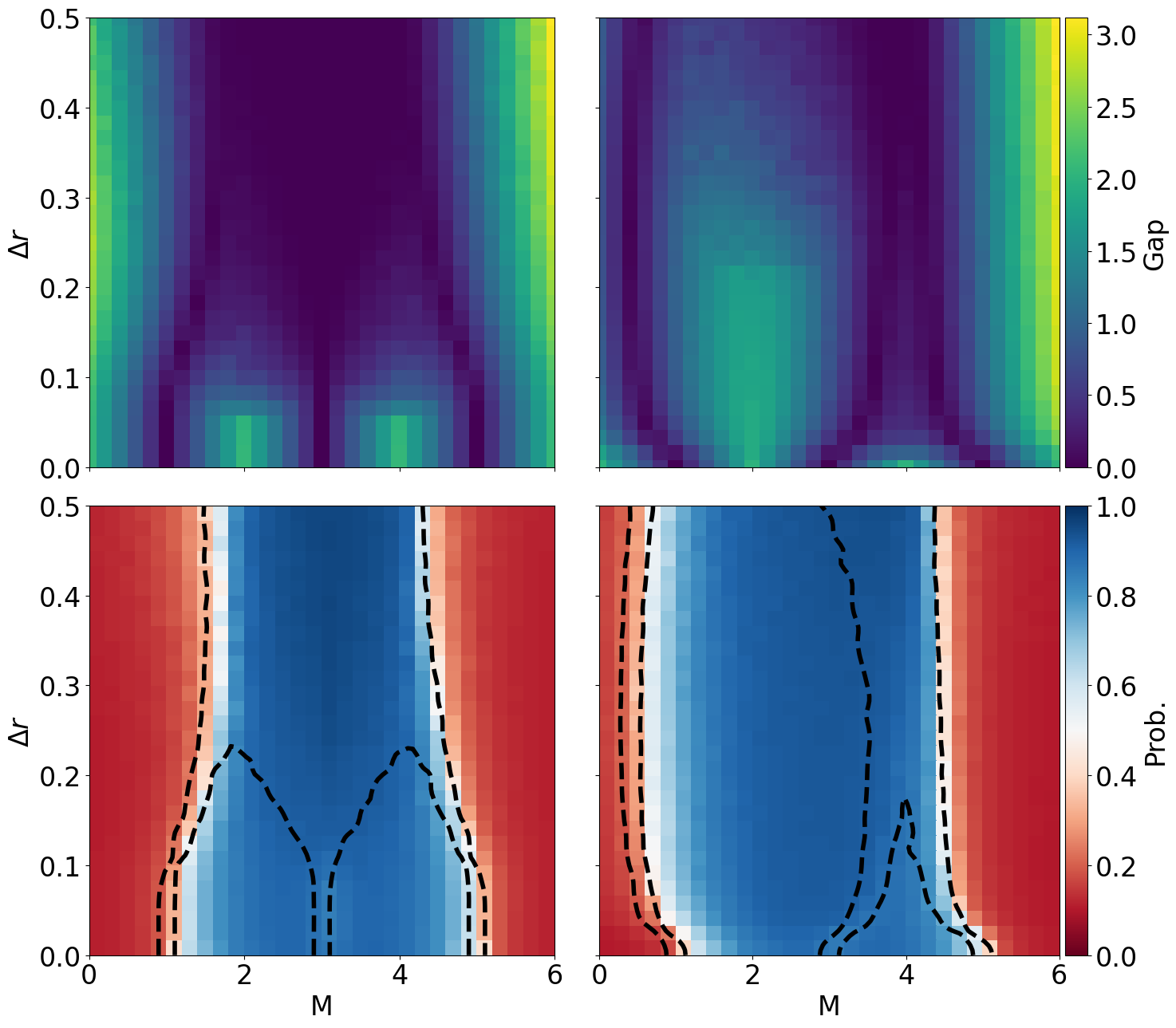}
    \caption{(Top) Gap diagrams for the Wilson-Fermion model on a square lattice as a function of the mass parameter $M$ and the displacement $\Delta r$ of the atoms, for two different cutoff distances. (Bottom) Topological phase diagrams predicted by the ANN, in terms of the outputted probability. Black lines correspond to contour lines
    from the gap diagram for $0.1$ eV. (Left) $R_c=1.1$, (Right) $R_c=1.4$. The cell size is $N=30$ (30 unit cells in each direction).}
    \label{wilson_gap}
\end{figure}

\textit{Amorphous solid}---
The specific disorder model we use is set in the following way.
Given a maximum displacement value $\Delta r$, which we take as the disorder parameter, we define the following random variables:
$$    r \sim U(0, \Delta r), \
 \theta \sim U(0, \pi/2), \
 \phi \sim  U(0, 2\pi)
$$
where $U(a,b)$ denotes a uniform distribution between $a, b$, $a < b$. Note that out-of-plane displacements are allowed.
Out of one sample of these variables, we generate a displacement vector given by 
$\Delta \textbf{r} = r(\sin\theta\sin\phi, \sin\theta\cos\phi, cos\theta)$, so the final position of each atom in the supercell is $\textbf{r}_i=\textbf{r}^0_i + \Delta \textbf{r}$,
where $\textbf{r}^0_i$ denotes the position if the lattice were crystalline. As we increase the value of  $\Delta r$ , the lattice becomes more disordered until long-range order is lost. 
The topological phase diagram will be obtained as a function of the maximum displacement $\Delta r$ and the mass parameter $M$.
In the following we will work with one supercell only, imposing periodic boundary conditions.

\begin{figure}[b]
    \centering
    \includegraphics[width=1\columnwidth]{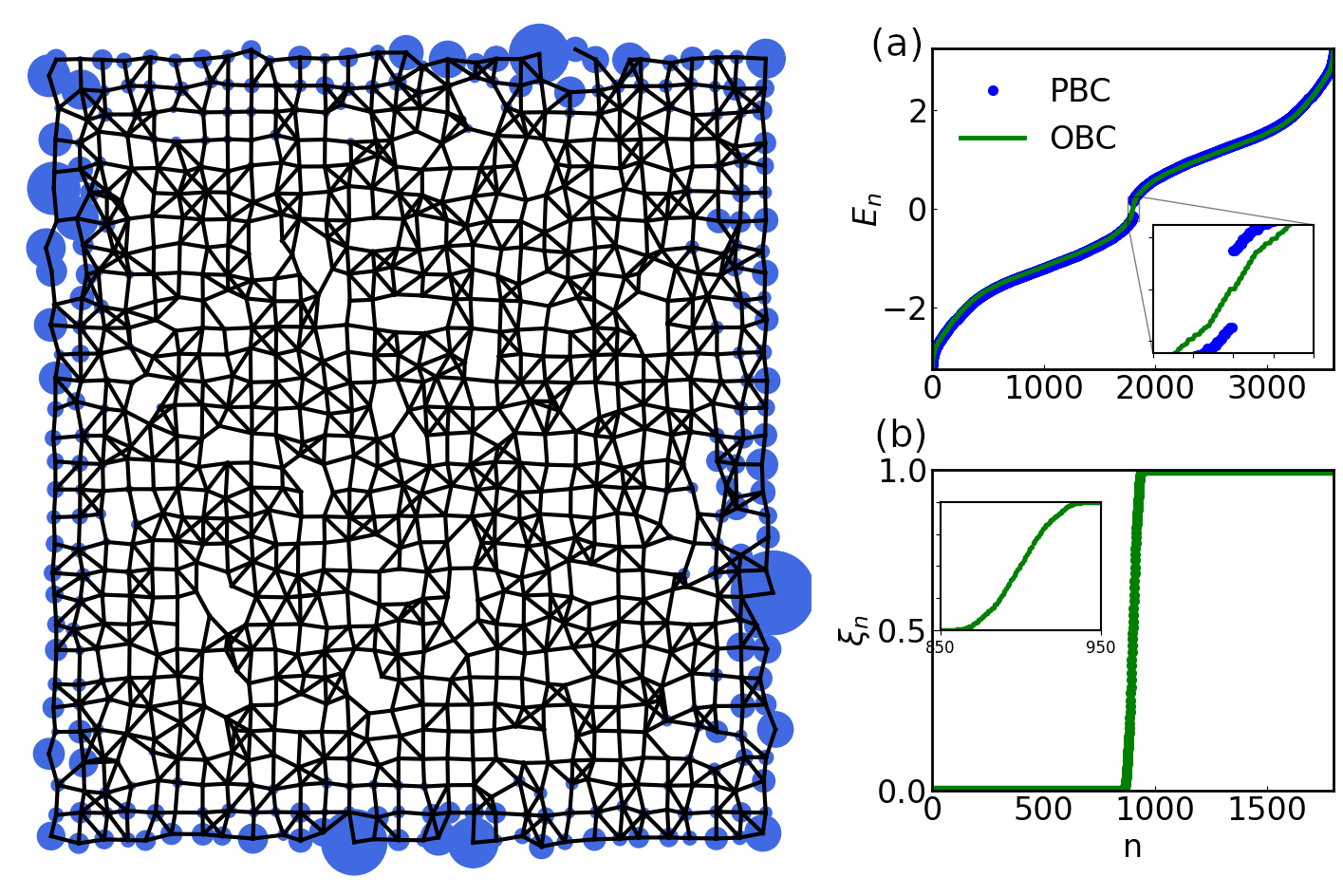}
    \caption{(a) Edge state, (b) Energy spectrum for both open and periodic boundary conditions and (c) Entanglement spectrum for $M=3$, $R_c=1.4$, $\Delta r = 0.5$.}
    \label{edge_state_r14}
\end{figure}

\begin{figure}[h]
    \centering
    \includegraphics[width=1\columnwidth]{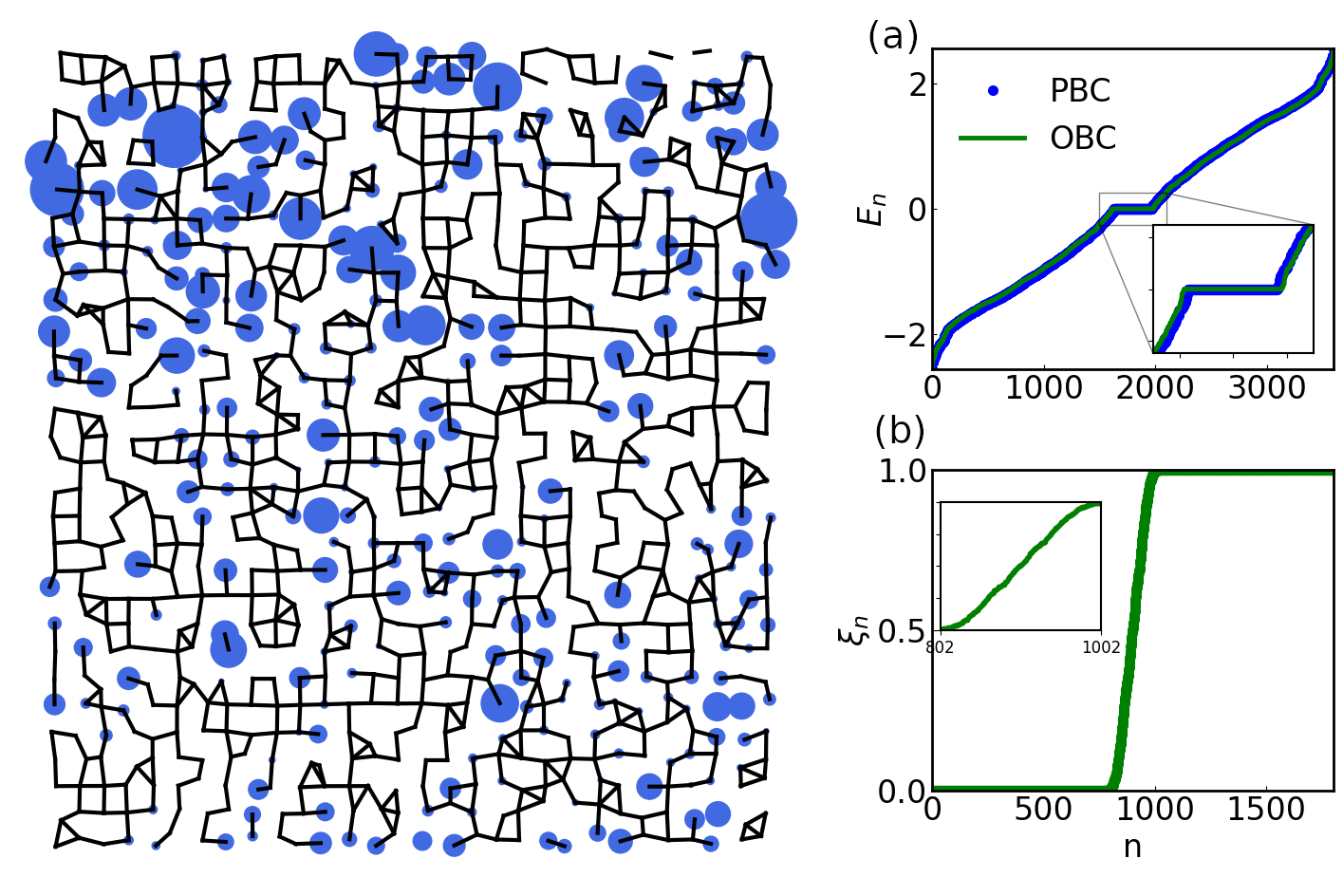}
    \caption{(a) Edge state, (b) energy spectrum for both OBC and PBC and (c) entanglement spectrum for $M=3$, $R_c=1.1$, $\Delta r = 0.5$.}
    \label{edge_state_r11}
\end{figure}

To obtain the phase diagrams, first we have to generate data, both for training and for prediction. We compute the entanglement spectrum for all combinations of $M$ and $\Delta r$ [see Figs. \ref{edge_state_r14}(c) and \ref{edge_state_r11}(c) and Suppl. Material for examples]. The training set will be given by the points corresponding to zero or very low disorder for $R_c=1.1$. As long as the functional form of the Hamiltonian remains the same, we expect the neural network to be valid even if it has not seen data from the system with $R_c=1.4$. 

\begin{figure}[t]
    \centering
    \includegraphics[width=1\columnwidth]{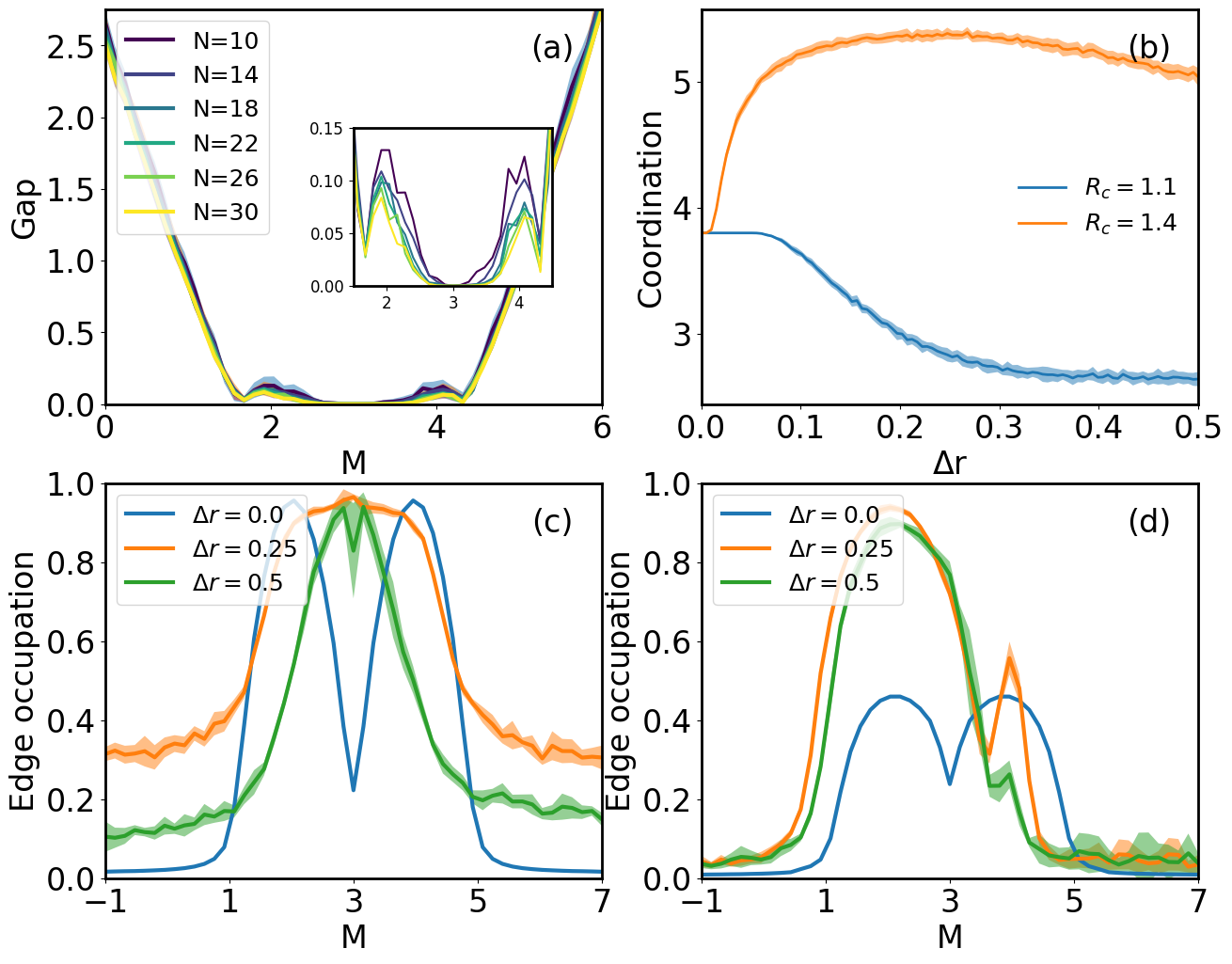}
    \caption{(a) Gap for $\Delta r = 0.25$ and different system sizes. (b) Coordination number of the solid for different cutoff distances as the disorder increases. (c) Average edge occupation of the lowest 21 eigenstates in energy for $R_c=1.1$ (d) Same as (c) for $R_c=1.4$.}
    \label{wilson_characterization}
\end{figure}

In Fig. \ref{wilson_gap} we show the topological phase diagrams obtained with the trained model for the largest systems studied (lower panels). We also show the gap (upper panels), which shows a weak dependence with the system  size [see Fig. \ref{wilson_characterization}(a)]. Here, we choose to plot the output of the ANN as an estimator of the probability of being in the topological phase. As commonly accepted, for probabilities higher than $0.5$ the system is considered to be topological. 
The model predicts the existence of topological states even in regions where the gap has vanished.

To verify the predictions of the ANN, we look for edge states near the Fermi energy where the ANN predicts non-trivial topology for disorder values outside the training set. In Fig. \ref{edge_state_r14}(a) we show an edge state obtained from one instance of the model near zero energy for $R_c=1.4$ and disorder $\Delta r = 0.5$. The gapful spectrum for this particular case is shown in Fig. \ref{edge_state_r14}(b) for periodic and open boundary conditions. We see a standard edge state in the sense that it is delocalized around the edge of the solid, as it would be expected. 
If we take a look at the bonds between atoms, the crystal has a high percolation due to more bonds appearing as we increase disorder.
However, for the smaller cutoff distance $R_c=1.1$, the solid starts to break, as shown by the diminishing coordination number in Fig. \ref{wilson_characterization}(b). This means that there are less paths available for a state to spread along, or equivalently that it has a lower percolation. If we now take a look at some edge state in the regions predicted to be non-trivial by the ANN, as in Fig. \ref{edge_state_r11}(a), we see that the occupation is not that we would have expected for an edge state, that is, around the borders of the solid. Still, looking closely we see that the electronic density appears mainly at the end of chains, which is the behaviour we would expect for one-dimensional topological systems. This clearly indicates that the system has undergone a transition from 2D to quasi-1D as it becomes increasingly disordered (due to the imposed cutoff between neighbours), while keeping a non-trivial topological nature. In this case, as Fig. \ref{edge_state_r11}(b) shows, there is no gap.

Finally, we can quantify the edge character in the transition from trivial to topological by looking at the average edge localization of  eigenstates near zero energy as a function of $M$, as shown in Fig. \ref{wilson_characterization}(c) and (d). As we approach the boundary between phases predicted by the ANN, there is a drastic change in edge localization, which is indicative of the phase transition.

\begin{figure}[h]
    \centering
    \includegraphics[width=1\columnwidth]{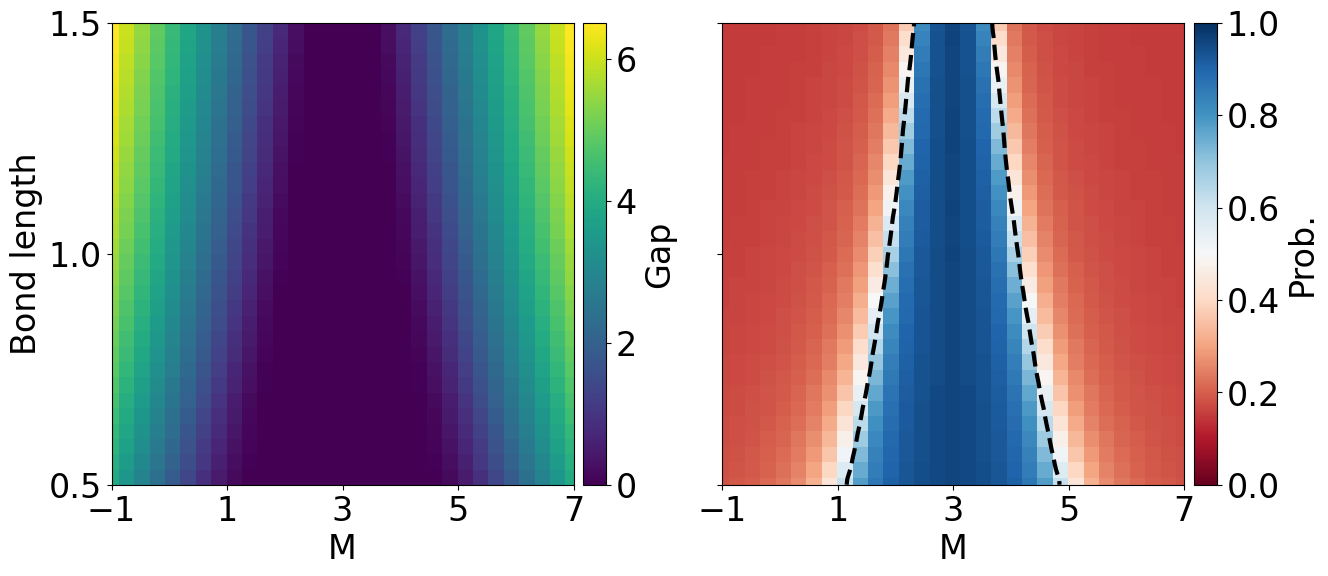}
    \caption{(Left) Gap diagram for the Bethe lattice with $z=3$ and depth$=8$. (Right) Topological phase diagram for the same model as predicted by the ANN trained with the Wilson-fermion model. Black lines correspond to contour lines from the gap diagram for $0.1$ eV.}
    \label{gap_bethe}
\end{figure}

\begin{figure}[t]
    \centering
    \includegraphics[width=1\columnwidth]{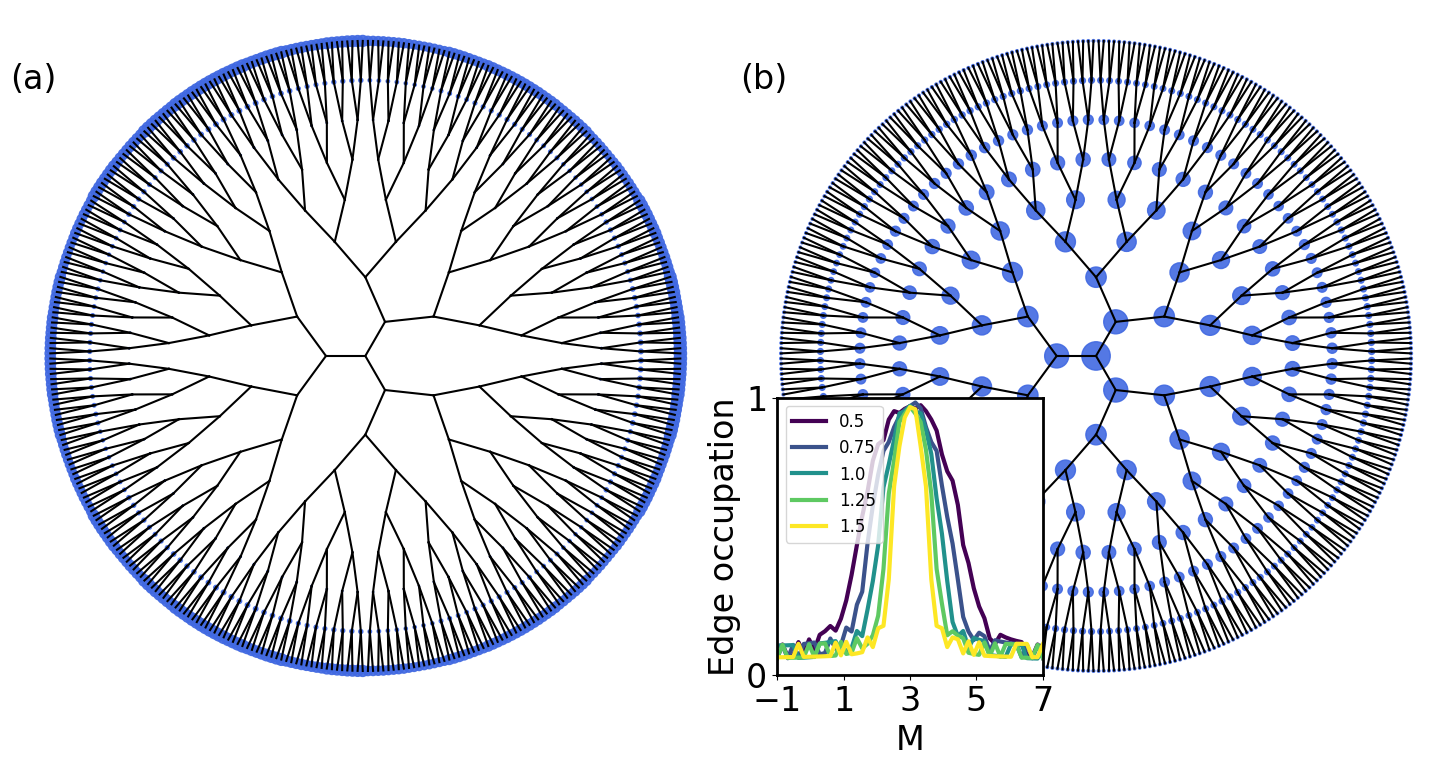}
    \caption{(a) Edge state for $M=2.5$, $l=0.7$. (b) Lowest absolute energy eigenstate for $M=1$, $l=1.4$. Inset: Edge occupation as a function of $M$ for different bond lengths $l$}
    \label{bethe_states}
\end{figure}

\textit{Bethe lattice}---
We consider next a different type of system without translational symmetry, namely a Bethe lattice, which is a type of fractal lattice where the $\mathbb{Z}_2$ invariant cannot be computed by standard means. Since the underlying model Hamiltonian is the same, we expect the previously trained neural network to predict the topological phase diagram of this system as well. The Bette lattice is defined by a coordination number $z$, which specifies the number of neighbours each atom has. Then, starting from a central node, the number of nodes in each consecutive layer is given by:
\begin{equation}
    N_k = z(z-1)^{k-1},\ k>0
\end{equation}
where $k$ denotes the $k$-th layer, e.g. in layer 1 there are three nodes. 
It is important to mention that, mathematically speaking, the Bethe lattice is realized if the above equation is fulfilled. For us, however, the specific arrangement of the atoms is relevant since the generalized Wilson model depends explicitly on the angles between the atoms. Thus, we arrange the atoms of each layer such that the angular spacing between them is uniform. Also, the distance between every connected pair of atoms is fixed. These two constraints, with the coordination number, reproduce the lattice shown in Fig.\ref{bethe_states}.

With the model established, we proceed in a analogous fashion as with the amorphous lattice. To obtain an equivalent topological phase diagram, first we must choose some parameters that define the parameter space. As opposed to the amorphous model, for the Bethe lattice there is not a disorder parameter since the lattice is fixed. Instead, we choose the bond length. The bond length affects the hopping amplitude, effectively changing the electronic structure. The corresponding gap diagram is shown in Fig. \ref{gap_bethe}(a). 

The entanglement spectrum for different combinations of the mass and bond length parameters is fed into the neural network, which predicts the phase diagram shown in Fig. \ref{gap_bethe}(b). In this case, the whole topological region has a negligible gap. To verify that the neural network is predicting correctly the different phases, we may represent eigenstates near the Fermi energy. In Fig. \ref{bethe_states}(a) we see how an edge state appears, similar to the ones in the amorphous system for $R_c=1.1$ (note that the probability density is located mainly at the end of the different branches, which in this case happens to be also the outermost atoms). For comparison, a trivial state is also shown in Fig. \ref{bethe_states}(b), as well as the average edge occupation for several eigenstates close to the Fermi level. In all cases the results are consistent with the diagram predicted by the ANN.
\\

\textit{Conclusions}---
We have shown that it is possible to use the entanglement spectrum of a non-translationally invariant system to train a neural network to predict topological and trivial phases. We have applied it, in particular, to the case of a disordered, even amorphous lattice and to a Bethe lattice with an underlying Wilson-Dirac fermion model. This method reduces the computational time for the calculation of the invariant once we have an ANN already trained, as opposed to using, e.g., the Wilson loop technique. More importantly, it can be used with gapless systems where no other method is currently available.
We expect that this method will allow to explore realistic models of disordered topological insulators, such as alloys as a function of the composition, topological metals and disorder-induced phase transitions in general.


\begin{acknowledgments}
The authors acknowledge financial support from Spanish MICIN through Grant No. PID2019- 109539GB-C43,  the María de Maeztu Program for Units of Excellence in R\&D (Grant No. CEX2018-000805-M), the Comunidad Autónoma de Madrid through the Nanomag COST-CM Program (Grant No. S2018/NMT-4321), the Generalitat Valenciana through Programa Prometeo/2021/01, the Centro de Computación Científica of the Universidad Autónoma de Madrid, and the computer resources of the Red Española de Supercomputación.
\end{acknowledgments}

\bibliographystyle{apsrev4-1}
\bibliography{biblio}

\end{document}